\documentstyle[12pt]{article}

\def\PL #1 #2 #3 {{\it Phys. Lett.} {\bf#1} (#3) #2}
\def\NP #1 #2 #3 {{\it Nucl. Phys.} {\bf#1} (#3) #2}
\def\ZP #1 #2 #3 {{\it Z. Phys.} {\bf#1} (#3) #2}
\def\PRL #1 #2 #3 {{\it Phys. Rev. Lett.} {\bf #1} (#3) #2}
\def\PR #1 #2 #3 {{\it Phys. Rev.} {\bf#1} (#3) #2}
\def\MPL #1 #2 #3 {{\it Mod. Phys. Lett.} {\bf#1} (#3) #2}
\def\RMP #1 #2 #3 {{\it Rev.~Mod. Phys.} {\bf#1} (#3) #2}
\def\lam{\ifm \lambda \else $\lambda$ \fi}
\def\ifm{\ifmmode}

\newcommand{\beq}{\begin{equation}}
\newcommand{\eeq}{\end{equation}}
\newcommand{\beqn}{\begin{eqnarray}}
\newcommand{\eeqn}{\end{eqnarray}}
\newcommand{\beqs}{\begin{eqnarray*}}
\newcommand{\eeqs}{\end{eqnarray*}}

 1
 1
 1

\begin{document}
\thispagestyle{empty}
\vglue -0.5cm
\hfill{FERMILAB-Pub-98/082-T}
\vskip 0.1cm
\hfill{March 1998}
\vskip 1cm
\begin{center}
{\large Implications of Hadron Collider Observables}
\vglue .4cm
{\large on Parton Distribution Function Uncertainties}
\vglue 1.4cm
\begin{sc}
Walter T. Giele and Stephane Keller\\
\vglue 0.2cm
\end{sc}
{\it Fermilab, MS 106\\
Batavia, IL 60510, USA}
\end{center}

\vglue 2cm
\begin{abstract}
\par \vskip .1in \noindent

Standard parton distribution function sets do not have rigorously
quantified uncertainties.  In recent years it has become apparent that
these uncertainties play an important role in the interpretation of
hadron collider data.  In this paper,
using the framework of statistical inference, we illustrate a
technique that can be used to efficiently propagate the uncertainties
to new observables, assess the compatibility of new data with an initial
fit, and, in case the compatibility is good, include the new data in
the fit.
\end{abstract}

\newpage
\setcounter{page}{1}
\section{Introduction}
Current standard sets of Parton Distribution Function (PDF) do not
include uncertainties~\cite{r1}.  In practice, as long as the PDF's
are used to calculate observables that themselves have large
experimental uncertainties this shortcoming is obviously not a
problem.  In the past the precision of the hadron collider data was
such that there was no ostensible need for the PDF uncertainties, as
was testified by the good agreement between the theory and measurements.
However, the need for PDF uncertainties became apparent with the
measurement of the one jet inclusive transverse energy at the
Tevatron~\cite{r2}.  At large transverse jet energies the data was
significantly above the theoretical prediction, a possible signal for
new physics.  The deviation was ultimately ``fixed'' by changing the
PDF's in such a manner that they still were consistent with
the observables used to determine the PDF~\cite{r3}.  
This is a reflection of the significant PDF
uncertainties for this observable.  Knowing the uncertainties on the
PDF's would have cleared the situation immediately.  Note that once
the data is used in the PDF fit, it can not be used for other
purposes.  Specifically, setting limits on possible physics beyond the
Standard Model.  In that case, one should fit the PDF's and the new
physics simultaneously.  The technique
presented in this paper is well suited for this sort of problem.

The spread between different sets of PDF's is often associated with
PDF uncertainties. Currently, this is what is used for the
determination of the PDF uncertainty on the $W$-boson mass at the
Tevatron.  It is not possible to argue that this spread is an accurate
representation of all experimental and theoretical PDF
uncertainties.  For the next planned high luminosity run at Fermilab,
assuming an integrated luminosity of $2\ fb^{-1}$,
the expected 40 MeV uncertainty on
the $W$-boson mass is dominated by a 30 MeV production model
uncertainty.
The latter uncertainty itself is dominated by the PDF uncertainty,
estimated to be 25 MeV~\cite{r4}.  This determination of the PDF
uncertainty is currently nothing more than an educated guess.  It is
made by ruling out existing PDF's using the lepton charge asymmetry in
$W$-boson decay events.  The spread of the remaining PDF's determines the
uncertainty on the extracted $W$-boson mass.  Because the PDF
uncertainty seems to be the dominant source of uncertainty in
the determination of the $W$-boson mass, such a procedure must be replaced
by a more rigorous quantitative approach.
The method described in this paper is well suited for this purpose.

In this paper, using the framework of statistical
inference~\cite{r5a,r5b}, we illustrate a method that can be used
for many purposes.  First of all, it is easy to propagate the PDF
uncertainties to a new observable without the need to calculate the
derivative of the observable with respect to the different PDF
parameters. Secondly, it is straightforward to assess the
compatibility of new data with the current fit and determine whether
the new data should be included in the fit.  Finally, the new data can
be included in the fit without redoing the whole fit.

This method is significantly different from more traditional
approaches to fit the PDF's to the data. It
is very flexible and beside solving the problems already mentioned, it
offers additional advantages.  First, the experimental uncertainties
and the probability density distributions for the fitted parameters do
not have to be Gaussian distributed.  However, such a generalization
would require a significant increase in computer resources.  Second,
once a fit has been made to all the data sets, a specific data set can
be easily excluded from the fit.  Such an option is important in order
to be able to investigate the effect of the different data sets. This
is particularly useful in the case of incompatible new data.  In that
case one can easily investigate the origin of the incompatibility.
Finally, because it is not necessary to redo a global fit in order to
include a new data set, experimenters can include their own new data
into the PDF's during the analysis phase.

The outline for the rest of the paper is as follows.  In
Sec.~\ref{sect:method}, we describe the inference method. The
flexibility and simplicity of the method is illustrated in
Sec.~\ref{sect:expanding}, by applying it to the CDF one jet inclusive
transverse jet energy distribution~\cite{r2} and the CDF lepton charge 
asymmetry data~\cite{r7}. In Sec.~\ref{sect:conclusions} we
draw our conclusions and outline future improvements and extensions to
our method.

\section{The Method of Inference}
\label{sect:method}

Statistical inference requires an initial probability density
distribution for the PDF parameters.  This initial distribution can be
rather arbitrary, in particular it can be solely based on theoretical
considerations.  Once enough experimental data are used to constrain
the probability density distribution of the parameters the initial
choices become irrelevant~\footnote{The standard PDF sets of
Ref.~\cite{r1} basically assume that the initial probability density
distribution for the parameters is uniform.}.  Obviously, the
initial choice does play a role at intermediate stages.  The initial
distribution can also be the result of a former fit to other data.
The data that we will use later in this paper do not constrain the
PDF's enough by themselves to consider using an initial distribution
based only on theory.  The final answer would depend too much on our
initial guess.  We therefore decided to use the results of
Ref.~\cite{r6}.  In this work the probability density distribution was
assumed to be Gaussian distributed and was constrained using Deep
Inelastic Scattering (DIS) data.  All the experimental uncertainties,
including correlations, were included in the fit, but no theoretical
uncertainties were considered.  The fact that no Tevatron data were
used allows us to illustrate the method with 
Tevatron data~\footnote{Recent PDF sets have also included the
Tevatron data that we will use, but none of these sets included
uncertainties.}.  We briefly summarize Ref.~\cite{r6} in the appendix.

In Sec.~2.1 we explain the propagation of the uncertainty to new
observables. Sec.~2.2 shows how the compatibility of new data with the
PDF can be estimated.  Finally, in Sec.~2.3 we demonstrate how the
effect of new data can be included in the PDF's by updating the
probability density distribution of the PDF parameters.

\subsection{Propagation of the uncertainty}

We now assume that the PDF's are parametrized at an initial
factorization scale $Q_0$, with $N_{par}$ parameters,
$\{\lambda\}\equiv\lambda_1, \lambda_2, \ldots, \lambda_{N_{par}}$ and
that the probability density distribution is given by
$P_{init}(\lambda)$.  Note that $P_{init}(\lambda)$
does not have to be a Gaussian distribution.

By definition $P_{init}(\lambda)$ is normalized to unity,
\begin{equation}
\int_V P_{init}(\lambda) d\lambda =1\ ,
\end{equation}
where the integration is performed over the full multi-dimensional
parameter space and $d\lambda\equiv \prod_{i=1}^{N_{par}} d\lambda_i$.
To calculate the parameter space integrals we use a
Monte-Carlo (MC) integration approach with importance sampling.  We
generate $N_{pdf}$ random sets of parameters $\{\lam\}$ distributed
according to $P_{init}(\lam)$. This choice should minimize the MC
uncertainty for most of the integrals we are interested in.  For
reference we also generate one set at the central values of the
$\{\lam\}$, the $\mu_{\{\lam\}}$.  The number of parameter sets to be used
depends on the quality of the data.  The smaller the experimental
uncertainty is compared to the PDF uncertainty, the more PDF's we
need. We must ensure a sufficient fraction of PDF's span the region of
interest (i.e.  close to the data).  For the purposes of this paper,
we found that $N_{pdf}=100$ is adequate.  Clearly, to each of the
$N_{pdf}$ sets of parameters $\{\lam\}$ correspond a set of unique
PDF's.  Each of these PDF sets have to be evolved using the
Altarelli-Parisi evolution equations.  We used the CTEQ package to do
this evolution~\cite{CTEQevol}.

We now can evaluate any integral $I$ over the parameter space
as a finite sum~\cite{r5b}
\begin{eqnarray}
\label{eq:I}
I &=& \int_V f(\lam) P_{init}(\lam) d\lam  \nonumber \\
  &\approx& \frac{1}{N_{pdf}} \sum_{j=1}^{N_{pdf}} f(\lam^j) \\
  &\equiv& \langle f\rangle\ ,\nonumber
\end{eqnarray}
with $\lam^j$ is the $j$-th random set of $\{\lam\}$.  
The function $f$ represents an integrable function of the PDF parameters.
The uncertainty on the integral $I$ due to the MC integration is given by,
\begin{equation}
\label{eq:deltaI}
\delta I= \sqrt{\frac { \langle f^2\rangle - \langle f\rangle^2}{N_{pdf}}}\ .
\end{equation}

For any quantity, $x(\lambda)$, 
that depends on the PDF parameters $\{\lambda\}$
(for example an observable, one of the flavor PDF's
or for that matter one of the parameter itself),
the theory prediction is given by its average value,
$\mu_x$, and its uncertainty, 
$\sigma_x $~\footnote{If the uncertainty 
distribution is not Gaussian
the average and the standard deviation might not properly 
quantify the distribution.}:
\begin{eqnarray}
\label{eq:musig}
\mu_x & = & \int_V x(\lambda) P_{init}(\lam) d\lam 
\approx\frac{1}{N_{pdf}}\sum_{j=1}^{N_{pdf}} 
x\left(\lambda^j\right)
\nonumber \\ 
\sigma_x^2 & = & \int_V (x(\lam)-\mu_x)^2 P_{init}(\lam) d\lam
\approx\frac{1}{N_{pdf}}\sum_{j=1}^{N_{pdf}} 
\left(x\left(\lambda^j\right)-\mu_x\right)^2\ .
\end{eqnarray}
Note that $\mu_x$ is not necessarily equal to the value of $x(\lam)$
evaluated at the central value of the $\{\lam\}$. However, this is how
observables are evaluated if one has only access to PDF's without
uncertainties.

Given $y(\lambda)$, another quantity 
calculable from the $\{\lam\}$, the covariance
of $x(\lambda)$ and $y(\lambda)$ is given by the usual expression:
\begin{eqnarray}
\label{corunc}
{\rm C}_{xy}&=&\int 
\left(x(\lam)-\mu_x\right)\left(y(\lam)-\mu_y\right) P_{init}(\lam) d\lam
\nonumber \\
&\approx&\frac{1}{N_{pdf}}\sum_{j=1}^{N_{pdf}} 
\left(x\left(\lambda^j\right)-\mu_x\right)
\left(y\left(\lambda^j\right)-\mu_y\right)\ .
\end{eqnarray}
The correlation between $x(\lambda)$ and $y(\lambda)$ is given by
${\rm cor}_{xy} = {\rm C}_{xy}/ (\sigma_x\sigma_y)$.  For example,
this can be used to calculate the correlation between two experimental
observables, between an observable and one of the PDF parameters, or
between an observable and a specific flavor PDF at a fixed Bjorken-$x$.

Using Eq.~\ref{eq:deltaI}, the MC uncertainty on the average 
and (co)variance is given by
\begin{eqnarray}
\delta \mu_x &=& \frac{\sigma_x}{\sqrt{N_{pdf}}} \nonumber \\
\delta \sigma_x^2 &=& \sigma_x^2 \sqrt{ \frac{2}{N_{pdf}} } \\
\delta {\rm C}_{xy} &=& {\rm C}_{xy} \sqrt{ \frac{2}{N_{pdf}} }\ . 
\nonumber
\end{eqnarray}

The MC technique presented in this sub-section, gives a simple way to
propagate uncertainties to a new observable, without the need for
calculating the derivatives of the observable with respect to the
parameters.

\subsection{Compatibility of New Data}

We will assume that one or several 
new experiments, not used in the
determination of the initial probability 
density distribution, have measured a set
of $N_{obs}$ observables $\{x^e\}=x^e_1,x^e_2,\ldots,x^e_{N_{obs}}$.
The experimental uncertainties, including the systematic
uncertainties, are summarized by the $N_{obs}\times N_{obs}$
experimental covariance matrix $C^{exp}$.  Note that the correlations
between experiments are easily incorporated.  Here however, we have to
assume that the new experiments are not correlated with any of the
experiments used in the determination of $P_{init}$.  The probability
density distribution of $\{x^e\}$ is given by
\begin{eqnarray}
\label{eq:pxe}
P(x^e) &=& \int_V P(x^e|\lam) P_{init}(\lam) d\lam \\
       &\approx& \frac{1}{N_{pdf}}\sum_{j=1}^{N_{pdf}} P(x^e|\lam^j)\ ,
\nonumber
\end{eqnarray}
where $P(x^e|\lam)$ is the conditional probability density
distribution (often referred to as likelihood function).  This distribution 
quantifies the probability of measuring the specific set of
experimental values \{$x^e$\} given the set of PDF parameters
$\{\lam\}$.  In PDF sets without uncertainties, $P_{init}(\lam)$ is a
delta function and $P(x^e)=P(x^e|\lam)$.

Instead of dealing with the probability density distribution of
Eq.~\ref{eq:pxe}, one often quotes the confidence level to
determine the agreement between the data and the model.
The confidence level is defined as the
probability that a repeat of the given experiment(s) would
observe a worse agreement with the model.
The confidence level of \{$x^e$\} is given by
\begin{eqnarray}
{\rm CL}(x^e) &=& \int_V {\rm CL}(x^e|\lam) P_{init} (\lam)d\lam \\
              &\approx& \frac{1}{N_{pdf}}\sum_{j=1}^{N_{pdf}} 
              {\rm CL}(x^e|\lam)\ , 
\nonumber
\end{eqnarray}
where ${\rm CL}(x^e|\lam)$ is the confidence level of \{$x^e$\} given 
$\{\lam\}$. If ${\rm CL}(x^e)$ is larger than an agreed value,
the data are considered consistent with 
the PDF and can be included in
the fit. If it is smaller, the data are inconsistent and we have
to determine the source of discrepancy.  

For non-Gaussian uncertainties  the calculation of the confidence level might
be ambiguous.  In this paper we assume that the uncertainties are Gaussian.  
The conditional
probability density distribution and confidence level are then given by
\begin{eqnarray}
P(x^e|\lam)=P(\chi_{new}^2) 
&=& \frac{e^{-\frac{1}{2}\chi^2_{new}(\lambda)}}
    {\sqrt{(2\pi)^{N_{obs}}|C^{tot}|}} \\
{\rm CL}(x^e|\lam)={\rm CL}(\chi_{new}^2)  
&=& \int_{\chi_{new}^2}^\infty P(\chi^2)\,d\chi^2\ ,
\end{eqnarray} 
where 
\begin{equation}\label{chisqnew}
\chi^2_{new}(\lambda) = \sum_{k,l}^{N_{obs}} 
                      \left(x^e_k-x_k^{t}(\lambda)\right)
           M^{tot}_{kl}\left(x^e_l-x_l^{t}(\lambda)\right)\ ,
\end{equation}
is the chi-squared of the new data. The theory prediction
for the $k$-th experimental observable, $x_k^t(\lam)$, is calculated
using the PDF set given by the parameters $\{\lam\}$.  The matrix
$M^{tot}$ is the inverse of the total covariance matrix, $C^{tot}$,
which in turn is given by the sum of the experimental, $C^{exp}$,
and theoretical, $C^{theor}$, covariance matrix.  We assume
that there is no correlation between the experimental and theoretical
uncertainties.  
We will use a minimal value of 0.27\% on the confidence level, 
corresponding to a three
sigma deviation, as a measure of compatibility of the data with
the theory. If the new
data are consistent with the theory prediction then the 
maximum of the distribution
of the $\chi^2_{new}$ should be close to $N_{obs}$ (within the
expected $\sqrt{2 N_{obs}}$ uncertainty). The
standard deviation of $\chi^2_{new}$,
$\sigma_{\chi^2_{new}}$, tells us something about
the relative size of the PDF uncertainty compared to the size of the
data uncertainty.  The larger the value of $\sigma_{\chi^2_{new}}$ is
compared to $\sqrt{2 N_{obs}}$, the more the data will be useful in
constraining the PDF's.

Note that if there are several uncorrelated experiments, the total
$\chi^2_{new}$ is equal to the sum of the $\chi^2_{new}$ of the
individual experiments and the conditional probability is equal to the
product of the individual conditional probabilities.

\subsection{Effect of new data on the PDF's}

Once we have decided that the new data are compatible with the initial
PDF's, we can constrain the PDF's
further.  We do this within the formalism of statistical inference,
using Bayes theorem.  The idea is to update the probability density
distribution taking into account the new data.  This new probability
density distribution is in fact the conditional probability density
distribution for the $\{\lam\}$ considering the new data \{$x^e$\} and is
given directly by Bayes theorem
\begin{equation}
\label{eq:pnew}
P_{new}(\lam)=P(\lam|x^e)=\frac{P(x^e|\lam)\ P_{init}(\lam)}{P(x^e)}\ ,
\end{equation}
where $P(x^e)$, defined in Eq.~\ref{eq:pxe}, acts as a normalization
factor such that $P(\lam|x^e)$ is normalized to one.  Because
$P_{new}(\lam)$ is normalized to unity, we can replace $P(x^e|\lam)$
in Eq.~\ref{eq:pnew} simply by $e^{-\frac{\chi^2_{new}(\lam)}{2}}$.  This
factor acts as a new weight on each of the PDF's.  

We can now replace $P_{init}(\lam)$ by $P_{new}(\lam)$ in the
expression for the average, standard deviation and covariance given in
Sec.~2.1 and obtain predictions that include the effect of the new
data.  With the MC integration technique described before, these
quantities can be estimated by weighted sums over the $N_{pdf}$ PDF
sets
\begin{eqnarray}
\label{eq:newmusig}
\mu_x &\approx&\sum_{k=1}^{N_{pdf}} w_k 
x\left(\lambda^{(k)}\right) \nonumber \\
\sigma_x^2 &\approx&\sum_{k=1}^{N_{pdf}} w_k 
\left(x(\lambda^{(k)})-\mu_x\right)^2 \\
C_{xy}&\approx&\sum_{k=1}^{N_{pdf}} w_k \left(x(\lambda^{(k)}-\mu_x\right)
                                \left(y(\lambda^{(k)}-\mu_y\right)\ , \nonumber
\end{eqnarray}
where the weights are given by
\begin{equation}
\label{wgt}
w_k=\frac{   e^{- \frac{1}{2} \chi_{new}^2(\lambda^{k} )}}
{\sum_{l=1}^{N_{pdf}}
e^{-\frac{1}{2} \chi_{new}^2(\lambda^{l}) }}\ .
\end{equation}
Note that for the calculation of the Monte-Carlo uncertainty of the weighted
sums, the correlation between the numerator and denominator 
in Eq.~\ref{eq:newmusig} has to be taken into account properly.

Our strategy is very flexible.  Once the theory predictions 
$x^t_l(\lam)$ using the $N_{pdf}$ PDF sets
are known for each of the experiments ,
it is trivial to include or exclude the effect of one of the
experiments on the probability density distribution.  If the different
experiments are uncorrelated then all what is needed is the $\chi^2_{new}$
of each individual experiments for all the PDF sets.  In that case,
each experiment is compressed into $N_{pdf}$ $\chi^2_{new}$ values.

One other advantage is that all the needed $x^t_l (\lam)$ can be calculated
beforehand in a systematic manner, whereas standard chi-squared or maximum
likelihood fits require many evaluations of $x_k^{t}(\lambda)$ during
the fit as the parameters are changed in order to find the extremum.
These methods are not very flexible, as a new fit is required each
time an experiment is added or removed.

The new probability density distribution of the PDF parameters is
Gaussian if the following three conditions are met.  First, the
initial probability density distribution, $P_{init}(\lam)$, must be
Gaussian.  Second, all the uncertainties on the data points must be
Gaussian distributed (that includes systematic and theoretical
uncertainties).  Finally, the theory predictions, $x_l^{t}(\lam)$,
must be linear in $\{\lam\}$ in the region of interest.  This
last requirement is fulfilled once the PDF uncertainties are small enough.
For the studies in this paper all three requirements are fulfilled.  
The new probability density distribution
can therefore be characterized  by the average value of the
parameters and their covariance matrix, which 
can be calculated, together with their MC
integration uncertainty,  using Eq.~\ref{eq:newmusig}.  
Once the new values of the average and the covariance matrix 
have been calculated, a new set of PDF parameters can
be generated according to the new distribution and used to make further
prediction instead of using the initial set of PDF with the weights.

An alternative way to generate a PDF set distributed according to
$P_{new}(\lam)$ is to unweight the now weighted initial PDF set.  The
simplest way to unweight the PDF sets is to use a rejection algorithm.
That is, define $w_{max}$ as the largest of the $N_{pdf}$ weights
given in Eq.~\ref{wgt}.  Next generate for each PDF set a uniform
stochastic number, $r_k$, between zero and one. If the weight $w_k$ is
larger or equal to $r_k\times w_{max}$ we keep PDF set $k$,
otherwise it is discarded.  The surviving PDF's are now distributed
according to $P_{new}(\lam)$.  The number of surviving PDF's is on
average given by $N_{pdf}^{new}=1/w_{max}$.  We can now apply all the
techniques of the previous sub-sections, using the new unweighted PDF
set. The MC integration uncertainties are easily estimated using the
expected number of surviving PDF's.  In the extreme case that $w_{max}$
is close to one and only a few PDF survive the unweighting procedure,
the number of initial PDF's must be increased.  The other extreme
occurs when all the weights are approximately equal, i.e. $w_k\sim
1/N_{pdf}$.  In that case the new data puts hardly any additional
constraints on the PDF.

The $\chi^2_{new}$ is only used to calculate the weight of a particular
PDF, so that the new probability density distribution of
the PDF parameters can be determined.  
We do not perform a chi-squared fit.  However, if
the new probability density distribution of the parameters is Gaussian
distributed then our method is equivalent to a chi-squared fit.  In that
case the average value of the parameters correspond to the maximum of
the probability density distribution.  The minimum chi-squared can be
estimated (with MC uncertainties) from the average $\chi^2_{new}$
calculated with the new probability density distribution.  Indeed, by
definition this average must be equal to the minimum chi-squared,
$\chi^2_{min}$, plus the known number of parameter.  Note that the
variance of the $\chi^2_{new}$ must itself be equal to twice the number
of parameters.  To obtain the overall minimum chi-squared, the value of
the minimum chi-squared of the initial fit must be added to
$\chi^2_{min}$.  As long as the confidence level of the new data that were
included in the fit is sufficiently high, the overall minimum chi-squared
obtained is guaranteed to be in accordance with expectations~\footnote{We are
assuming that the initial $\chi^2_{min}$ was within expectations.}.

\section{Expanding the PDF sets}
\label{sect:expanding}

The viability of the method described in Sec.~2 is studied using
two CDF measurements. In Sec.~3.1 the one jet inclusive
transverse energy distribution is considered, while the lepton charge 
asymmetry in $W$-boson decay is examined
in Sec.~3.2.
The statistical, 
systematic, and theoretical uncertainties on
the observables will be taken into account.

\subsection{The one jet inclusive measurement}

The CDF results on the one jet inclusive transverse energy
distribution~\cite{r2} demonstrated the weakness of the current
standard PDF sets due to the absence of uncertainties on the PDF
parameters.

The observables are the inclusive jet cross section at different 
transverse energies~\footnote{To be more precise, 
the inclusive jet cross section in different bins of transverse
energy. In the numerical results presented here we take 
the finite binning effects into account.}, $E_T^i$
\begin{equation}
x_i=\frac{d\,\sigma}{d\, E_T}(E_T^i)\ .
\end{equation}  
We first have to construct the experimental covariance matrix,
$C^{exp}_{ij}$, using the information contained in Ref.~\cite{r2}.
The paper lists the statistical uncertainty at the different
experimental points, $\Delta_0 (E_T^i)$, together with eight
independent sources of systematic uncertainties, $\Delta_k (E_T^i)$.
Hence, the experimental measurements, $x_i^e$ are given by
\begin{equation}
\label{eq:xie}
x_i^e =  x_i^t (\lam) 
+ \eta_0^i \Delta_0 (E_T^i)
+ \sum_{k=1}^8 \eta_k \Delta_k (E_T^i)\ ,
\end{equation}
where as before, $x_i^t (\lam)$ is the theoretical 
prediction for the observable calculated 
with the set of parameters $\{\lam\}$.
The $\eta_0^i$ and $\eta_k$ are independent 
random variables normally distributed
with zero average and unit standard deviation.
Note that some of the systematic uncertainties given in Ref.~\cite{r2}
are asymmetric. In those cases we symmetrized the uncertainty using
the average deviation from zero. 
From Eq.~\ref{eq:xie} we can construct the experimental covariance matrix
\begin{equation}
C^{exp}_{ij}=\left(\Delta_0 (E_T^i)\right)^2 \delta_{ij}
         +\sum_{k=1}^8\Delta_k (E_T^i) \Delta_k (E_T^j)\ .
\end{equation}

We also need to estimate the theoretical uncertainty. 
In Eq.~\ref{eq:xie} no theoretical uncertainties were taken into account.
We consider two types
of uncertainties: the uncertainty due to the numerical Monte 
Carlo integration over the final state particle phase space, 
$\Delta_{MC}(E_T^i)$, and the 
renormalization/factorization scale, $\mu$, uncertainty, 
$\Delta_{\mu} (E_T^i)$.
The theoretical prediction in Eq.~\ref{eq:xie} must then be replaced by
\begin{equation}
x_i^t (\lam) \rightarrow \frac{d\,\sigma^{NLO}}{d\,E_T}
(E_T^i,\lam,\mu) 
+ \eta_{MC}^i \Delta_{MC}(E_T^i)
+ \eta_{\mu} \Delta_{\mu} (E_T^i)\ ,
\end{equation}
from which we can derive the theoretical covariance matrix
\begin{equation}
C^{theor}_{ij}=\left(\Delta_{MC}(E_T^i)\right)^2 \delta_{ij}
+\Delta_{\mu} (E_T^i) \Delta_{\mu} (E_T^j)\ .
\end{equation}
Here we assume that there is no bin to bin correlation in the MC
uncertainty.  On the other hand, we take the correlation of the scale
uncertainty fully into account.  Both $\Delta_{MC}$ and $\Delta_{\mu}$
are evaluated at the central values of the PDF parameters, assuming
that the variation is small.

We evaluate the scale uncertainty in a very straightforward manner.
As the central prediction the renormalization and factorization scale
are taken to be equal to half the transverse energy of the leading jet in the
event, $\mu=\frac{1}{2}E_T^{max}$.  To estimate the uncertainty we
make another theoretical prediction now choosing as a scale
$\mu=E_T^{max}$. The ``one-sigma'' uncertainty is defined as
\begin{equation}
\Delta_{\mu}(E_T) =
  \frac{d\,\sigma^{NLO}}{d\,E_T}(E_T,\mu_{\lam},\mu=\frac{1}{2}E_T^{max})
- \frac{d\,\sigma^{NLO}}{d\,E_T}(E_T,\mu_{\lam},\mu=E_T^{max})\ .
\end{equation}
As we will see later in this section the theoretical uncertainties 
are small compared to the other 
uncertainties.  Therefore this crude estimate suffices for 
the purposes of this paper. In the future a more detailed
study of the theoretical uncertainty is required.
The scale uncertainty is often associated with the theoretical
uncertainty due to the truncation of the perturbative series.
However, it is important to realize this is only a part of
the full theoretical uncertainty.

\begin{figure}\vspace{8cm}
\includegraphics{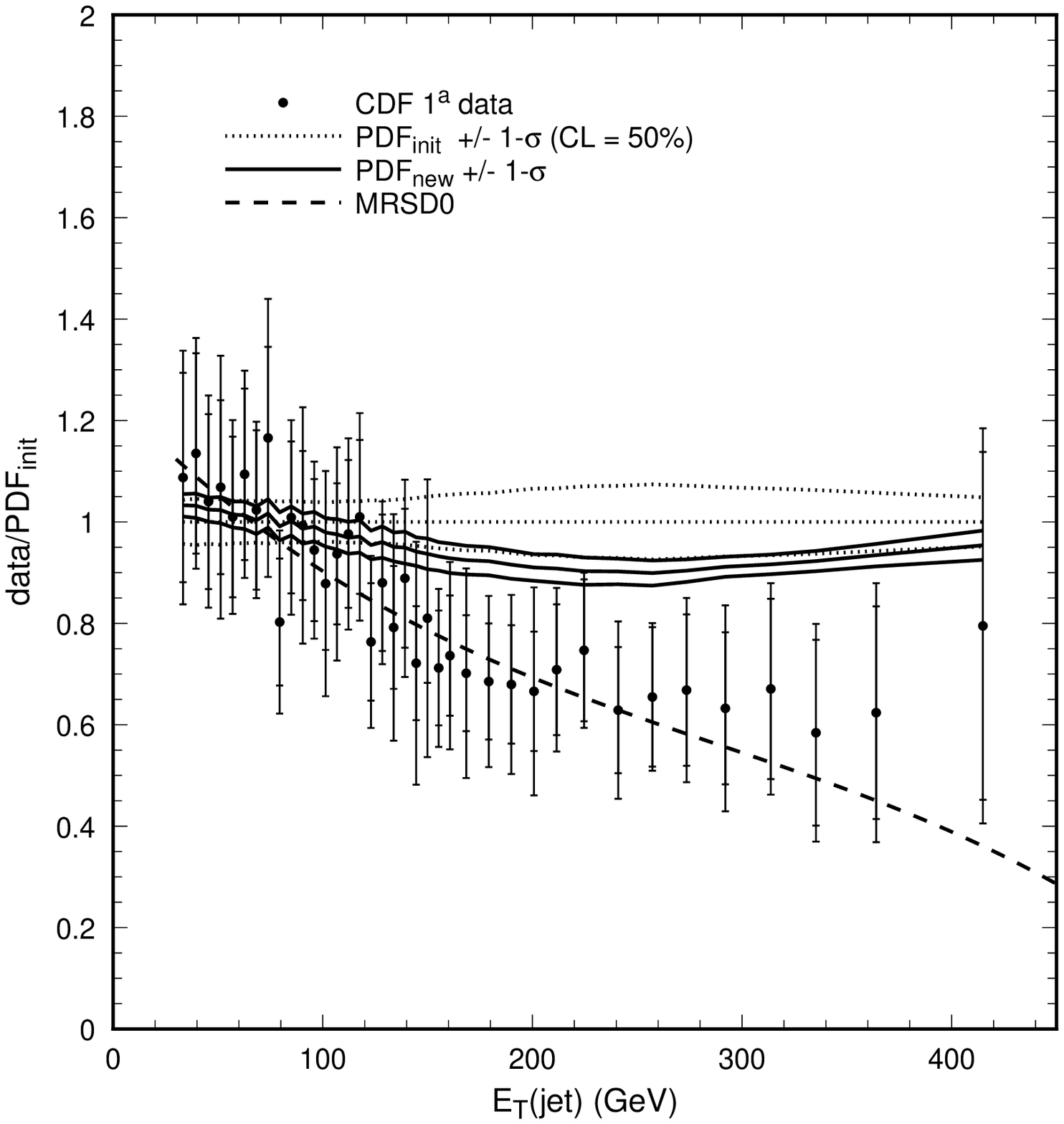}
\includegraphics{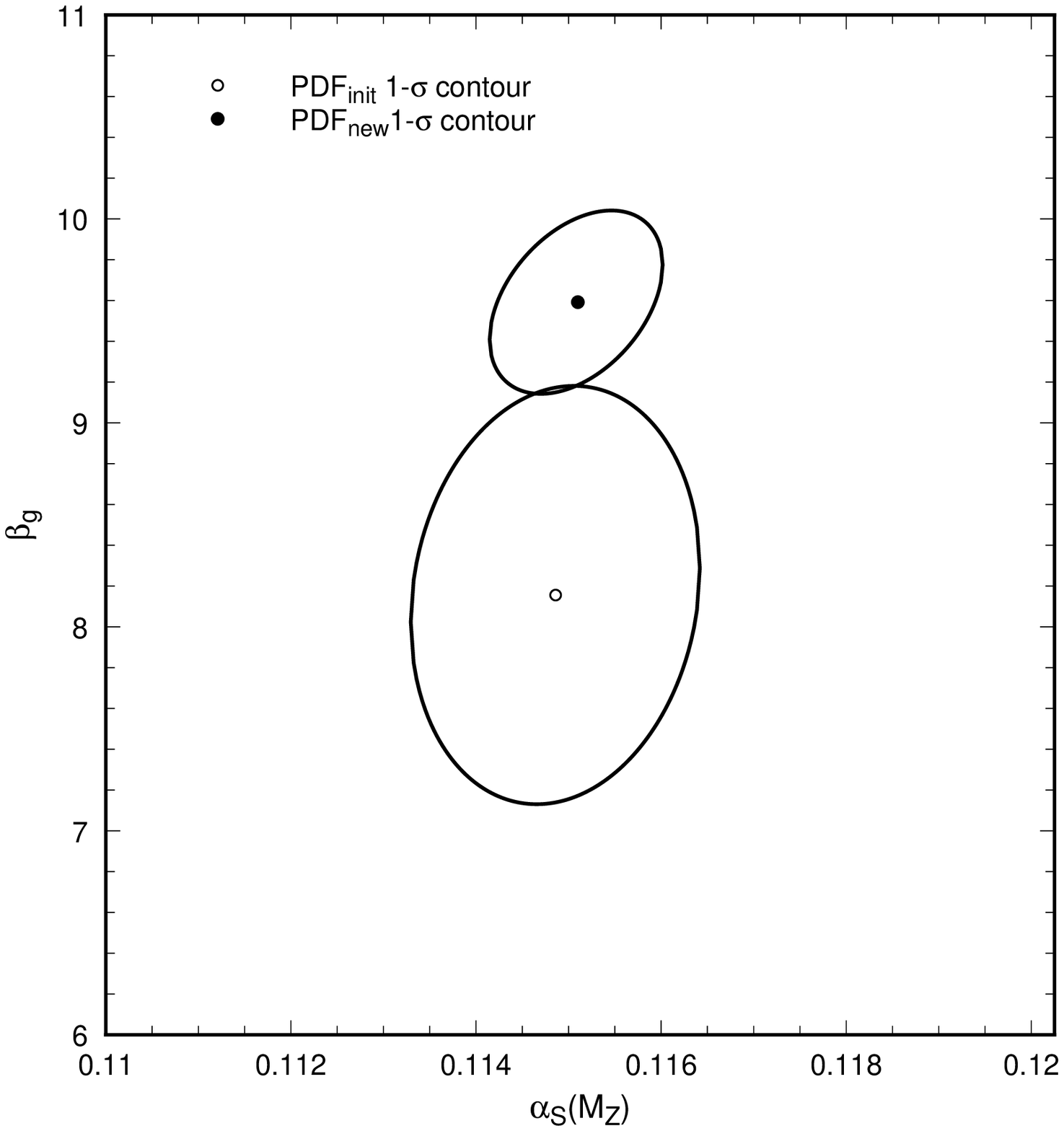}
\caption[]{ (a) Single inclusive jet cross section as a function of
the jet transverse energy.  The results are divided by the average
prediction calculated with the initial PDF's.  The data points are the
CDF run~1$^a$ results.  The dotted lines represent the initial one-sigma
PDF uncertainties.  The solid lines are the theory predictions
calculated with the new PDF's.  The inner (outer) error bars on the data
points are the diagonal entries of the experimental (total) covariance
matrix.  The dashed line is the prediction obtained with the MRSD0 PDF
set.  (b) The one-sigma correlation contour between the strong
coupling constant $\alpha_S(M_Z)$ and the $\beta$-parameter in the
gluon PDF ($\simeq x^\alpha(1-x)^\beta$ at the initial factorization
scale) calculated for both the initial and new PDF's.}
\label{fig3}
\end{figure}
In Fig.~\ref{fig3}$^a$ we present results for the single inclusive jet
cross section as a function of the jet transverse energy.  Both data
and theoretical predictions are divided by the average prediction of
the initial PDF's.  The NLO predictions are calculated using the
JETRAD prediction~\cite{jetrad}.  The inner (outer) error bar on the
experimental points represent the diagonal part of the experimental
(total) covariance matrix.  The dotted lines represent the initial
one-sigma PDF uncertainties.  The solid lines are the theory
predictions calculated with the new PDF's (i.e., the new probability density
distribution).  The plot is somewhat misleading because of the large
point-to-point correlation of the uncertainties.  The confidence level
of 50\% is very high, indicating a good agreement between the
prediction and the data.  

This leads us to the conclusion that the one jet inclusive transverse
energy distribution is statistically in agreement with the NLO
theoretical expectation based on the initial probability density
distribution of the PDF parameters. No indication of new physics is
present.  Note that the prediction using the initial PDF differs quite
a bit from the more traditional fits such as MRSD0, see the dashed
line in Fig.~\ref{fig3}$^a$.  Having no uncertainties on the
traditional fits it is hard to draw any quantitative conclusion from
this observation.  The larger value of the jet cross section 
calculated using the initial PDF set at high
transverse energies compared to MRSD0 was anticipated in
Ref.~\cite{r6} and can probably be traced back to the larger $d$ and
$u$ quark distribution at the reference scale $Q_0$ and moderate
$x\sim 0.2$.  This difference in turn was partially attributed to the
different way of treating target mass and Fermi motion corrections.

Given the confidence level of 50\% the one jet inclusive data can be
included in the fit. Using Eq.~\ref{chisqnew} we calculate for each
PDF set $k$ the corresponding $\chi^2_{new}(\lambda^k)$.  This gives
us the 100 weights $w_k$ (conditional probabilities) defined in
Eq.~\ref{wgt}.  Using Eq.~\ref{eq:newmusig}, we can calculate the
effects of including the CDF data into the fit.  The results are shown
in Figs.~\ref{fig3}$^a$ and~\ref{fig3}$^b$.  As can be seen in
Fig.~\ref{fig3}$^a$ the effect is that the central value is pulled
closer to the data and the PDF uncertainty is reduced substantially.
Two of the fourteen PDF parameters are affected the most.  As expected
these are the strong coupling constant $\alpha_S(M_Z)$ and the gluon
PDF coefficient $\beta$, which controls the high $x$ behavior (the
gluon PDF is proportional to $x^\alpha (1-x)^\beta$ at the initial
scale).  In Fig.~\ref{fig3}$^b$ we show the correlation between these
two parameters before and after the inclusion of the CDF
data. As can be seen the impact on $\beta$ is very significant.
Similarly, the uncertainty on $\alpha_s$ is reduced substantially and
the correlation between the two parameters is also changed.  This
indicates that the one jet inclusive transverse energy distribution in
itself has a major impact on the uncertainty of $\alpha_s$ and the
determination of the gluon PDF.  Note that we do not address the issue
of the parametrization uncertainty.  Other choices of how to
parameterize the initial PDF's will change the results.  To obtain a
value and uncertainty of $\alpha_S(M_Z)$ which is on the same footing
as the one obtained from $e^+e^-$-colliders, one needs to address this
issue.

\subsection{The lepton charge asymmetry measurement}

\begin{figure}\vspace{8cm}
\includegraphics{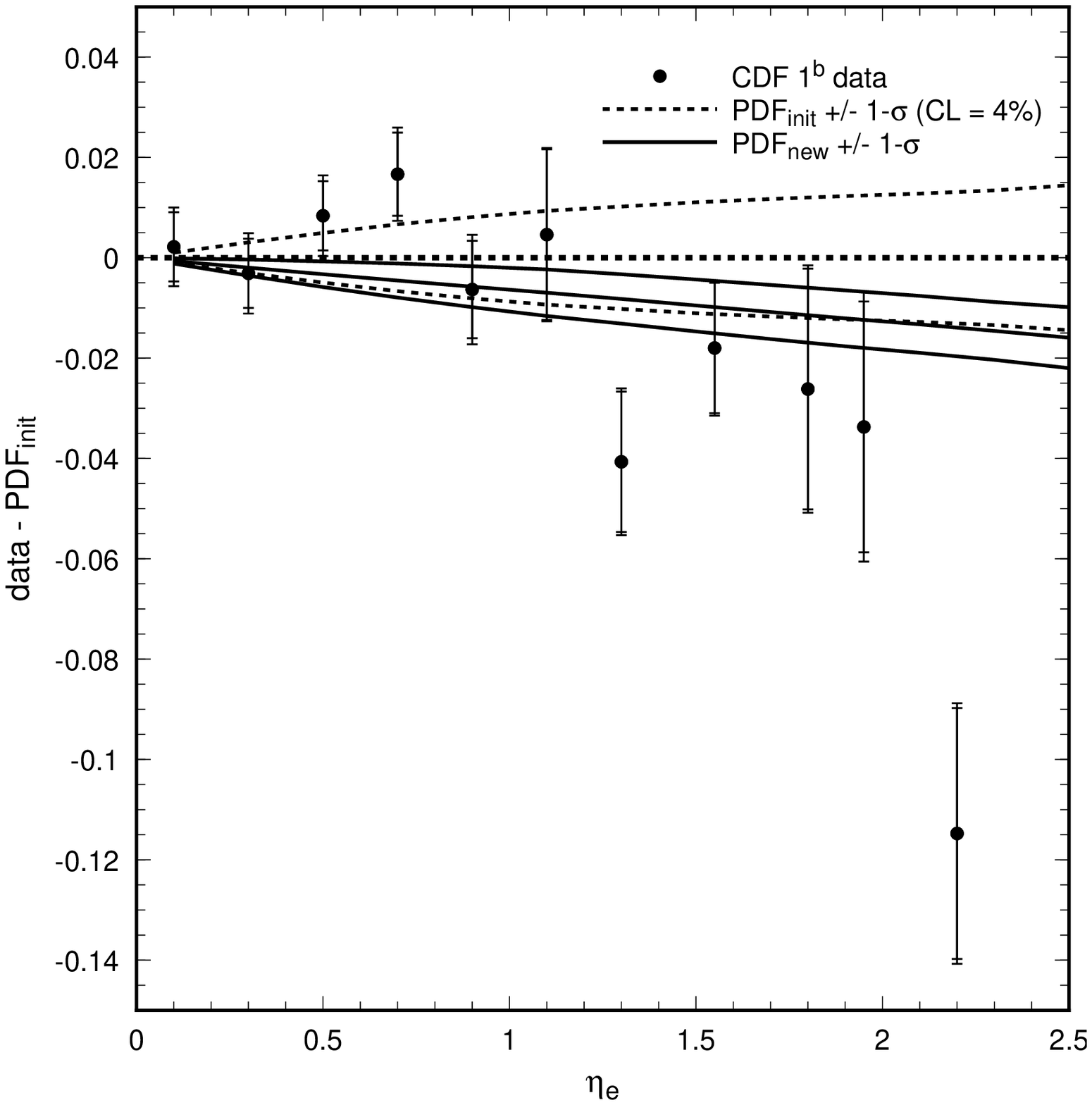}
\includegraphics{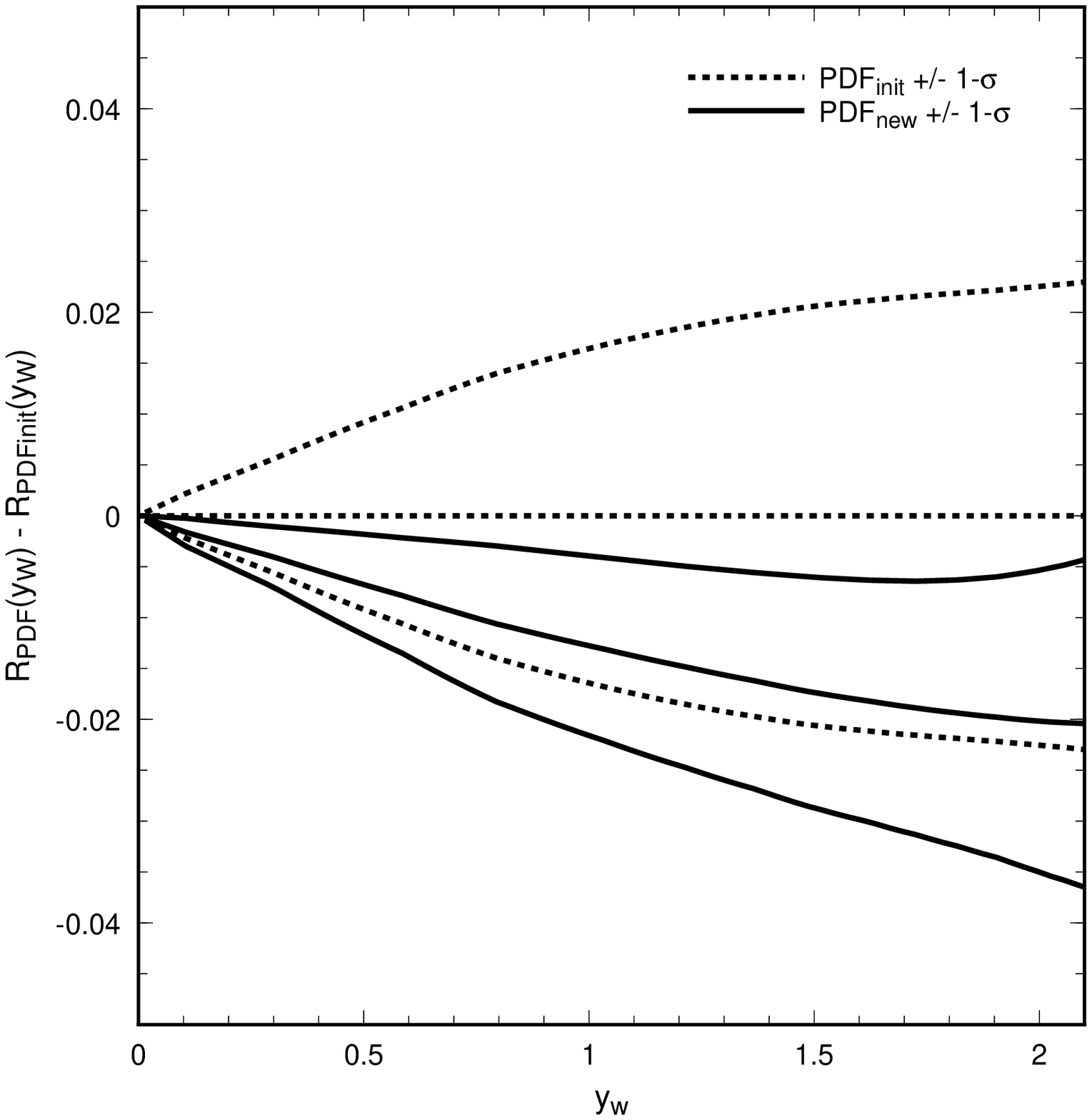}
\caption[]{ (a) The lepton charge asymmetry as a function of the
lepton pseudo-rapidity.  The results are normalized to the theory
prediction using the average value of the initial PDF's.  The data are
the CDF run~1$^b$ preliminary results.  The error bars, dotted and solid
lines have the same definition as in Fig.~1.  (b) The ratio $R(y_W)$
normalized as in (a) as a function of the $W$-boson rapidity.  The
dotted and solid lines are defined as in Fig.~1.}
\label{fig:asym}
\label{RyW}
\end{figure}

Our second example is the lepton charge asymmetry in $W$-boson decay
at the Tevatron.  As already explained, this observable is important
for the reduction of the PDF uncertainties in the $W$-boson mass
extraction at hadron colliders.  The asymmetry is given by
\begin{equation}\label{asym}
A(\eta_e)=\frac{\left(N^+ (\eta_e)-N^- (\eta_e)\right)}
{\left(N^+ (\eta_e)+N^- (\eta_e)\right)}\ ,
\end{equation}
where $N^+$ and $N^-$ are respectively
the number of positrons and electrons at the
pseudo-rapidity $\eta_e$.    

In Fig.~\ref{fig:asym}$^a$, we show the preliminary CDF data of 
run~1$^b$ (solid points) for the asymmetry, along with the NLO predictions
(dotted lines) including the PDF uncertainties, relative to the theory
average prediction using the initial PDF's.  For the NLO calculations
the DYRAD prediction~\cite{jetrad} was used.  The inner error bars on
the experimental points are the statistical uncertainties; the
systematic uncertainties are small and we can safely neglect them.
The outer error bars are the diagonal of the total covariance matrix.
In this case, the theoretical uncertainty is dominated by the phase space 
Monte Carlo integration uncertainty; we took its bin to
bin correlation into account.  Similar to the one jet inclusive
transverse energy case, the scale uncertainty is defined by the
difference between the theoretical prediction calculated using two
scales, $\mu=M_W$ and $\mu=2\times M_W$.

As is clear from Fig.~\ref{fig:asym}$^a$, there is a good agreement
between the data and the NLO prediction, except for the last
experimental point at the highest pseudo-rapidity.  The confidence
level including the last point is well below our threshold of 0.27\%.
In order to be able to include the data into the PDF fit we decided to
simply exclude this data point from our analysis.  Without the highest
pseudo-rapidity point we obtain a reasonable confidence level of 4\%.
It is not as good as in the single inclusive jet case even though the
plots appear to indicate otherwise. The reason for this is the absence
of significant point-to-point correlation for the charge asymmetry 
uncertainties.

We can now include the lepton charge asymmetry data into the fit by
updating the probability density distribution with Bayes theorem, as
described in the previous section.  In Fig.~\ref{fig:asym}$^a$ the
prediction obtained with the new probability density distribution are
shown by the solid lines.  As expected, the data are pulling the
theory down and reducing the PDF uncertainties.

It is difficult to correlate the change in the asymmetry to a change
in a particular PDF parameter.  On the other hand, it is well known
that the lepton asymmetry can be approximately related to the
following asymmetry of the ratio of up quark ($u$) and down quark
($d$) distribution function
\begin{equation}
R(y_W)= \frac{\frac{u(x_1)}{d(x_1)}-\frac{u(x_2)}{d(x_2)}}
{\frac{u(x_1)}{d(x_1)}+\frac{u(x_2)}{d(x_2)}}\ .
\end{equation}
The Bjorken-$x$ are given by
\begin{equation}
x_{1,2}=\frac{M_W}{\sqrt{s}} e^{\pm y_W}\ .
\end{equation}
where $M_W$ is the mass of the $W$-boson, ${\sqrt{s}}$ the center of
mass of the collider, and $y_W$ the $W$-boson rapidity.  The PDF's were
evaluated with the factorization scale equal to $M_W$.  
The ratio $R(y_W)$ is
approximately the $W$-boson asymmetry and obviously is sensitive to the
slope of the $u$/$d$ ratio.

In Fig.~\ref{RyW}$^b$ we show the ratio $R(y_W)$ calculated with both
the initial and the new probability density distributions.  As can be
seen, the change is very similar to the change experienced by the
lepton charge asymmetry itself.  The change in $R(y_W)$ can be traced
to a simultaneous increase in the anti-up quark distribution and
decrease in the anti-down quark distribution at low $x$.

\section{Conclusions ant Outlook}
\label{sect:conclusions}

Current standard sets of PDF do not include uncertainties.  It is
clear that we can not continue to discount them.  Already current
measurements at the Tevatron have highlighted the importance of these
uncertainties for the search of physics beyond the Standard Model.
Furthermore, the potential of future hadron colliders to measure
$\alpha_s(M_Z)$ and the $W$-boson mass is impressive, but can not be
disentangled from PDF uncertainties.  The physics at the LHC will also
undoubtedly require a good understanding of the PDF uncertainties.  On
a more general level, if we want to quantitatively test the framework
of perturbative QCD over a very large range of parton collision
energies the issue of PDF uncertainties can not be sidestepped.

In this paper we have illustrated a method, based on statistical
inference, that can be used to easily propagate uncertainties to new
observables, assess the compatibility of new data, and if the latter
is good to include the effect of the new data on the PDF without
having to redo the whole fit.  The method is versatile and modular: an
experiment can be included in or excluded from
 the new PDF fit without any
additional work.  The statistical and systematic uncertainties with
the full point-to-point correlation matrix can be included as well as
the theoretical uncertainties.  None of the uncertainties are required
to be Gaussian distributed.

One remaining problem is the uncertainty associated with the choice of
parametrization of the input PDF.  This is a difficult problem that
does not have a clear answer yet and will require a compromise
between the number of parameters and the smoothness of the PDF.  We
plan to address this question in another paper.  The next phase would
then be to obtain a large number of initial PDF's sets based on
theoretical consideration only, in the spirit of the inference method
and Bayes theorem.  The DIS and Tevatron data could then be used to
constraint the range of these PDF's resulting in a set of PDF's 
which would include both experimental and theoretical uncertainties.

\appendix
\section{Appendix: Input PDF}
\label{sect:input}

For our initial PDF parameter probability density distribution we use
the results of Ref.~\cite{r6}.  There a chi-squared fit was performed to
DIS data from BCDMS~\cite{bcdms}, NMC~\cite{nmc}, H1~\cite{h1} and
ZEUS~\cite{zeus}.  Both statistical uncertainties and experimental
systematic uncertainties with point-to-point correlations were
included in the fit, assuming Gaussian distributions.  However, {\it
no} theoretical uncertainties were considered. It is important to
include the correlation of the systematic uncertainties because the
latter usually dominate in DIS data.  Simply adding them in quadrature
to the statistical uncertainty would result in a overestimation of the
uncertainty.

A standard parametrization at $Q_0^2=9 \ {\rm GeV}^2$ is used with 14
(=$N_{par}$) parameters: $x d_v$, $x g$, $x \bar{d}$, $x \bar{u}$, and
$x s$ are parametrized using the
functional form $x^{\lambda_i} (1-x)^{\lambda_j}$
whereas $x u_v$ is parametrized as $x^{\lambda_i}
(1-x)^{\lambda_j} (1+\lambda_k x)$. Here $x$ is the
Bjorken-$x$. Parton number and momentum conservation 
constraints are imposed.  The
full covariance matrix of the parameters, $C^{init}$, is extracted at
the same time as the value of the parameters that minimize the
chi-squared.  The uncertainties on the parameters were assumed to be
Gaussian, such that the fitted values also correspond to the average
values of the parameters, $\mu_{\lam_i}$.  The probability density
distribution is then given by
\begin{equation}
\label{eq:chi2init}
P_{init}(\lam) = \frac{e^{-\frac{\chi^2_{init}(\lam)} {2} }}
{\sqrt{(2\pi)^{N_{par}}|C^{init}|}}\ ,
\end{equation}
where 
\begin{equation}
\label{chiinit}
\chi^2_{init}(\lam)=\sum_{ij}^{N_{par}}(\lambda_i-\mu_{\lambda_i})
               M^{init}_{ij}(\lambda_j-\mu_{\lambda_j})\ ,
\end{equation}
is the difference between the total chi-squared of the experimental data 
used in the fit and the minimum chi-squared (1256 for 1061 data points)
with the PDF's fixed by the set of parameters $\{\lam\}$.  
The matrix $M^{init}$ is the inverse of the covariance matrix
$C^{init}$.  The $|C^{init}|$ is the determinant of the covariance
matrix.  All the calculations were done in the $\overline{MS}$-scheme.

Comparison with MRS and CTEQ sets showed a good overall agreement with
a few exceptions.  One example is the difference in the gluon
distribution function at large values of $x$.  The CTEQ and MRS
distribution are somewhat above the result of Ref.~\cite{r6}.  This
difference was attributed to the fact that prompt photon data were
included in the CTEQ and MRS fits.  Note that the direct photon data
have large scale uncertainty, and it might be misleading to include
them in a fit without taking the theoretical uncertainty into account.
Also, it is important to keep in mind that it is misleading to compare
specific PDF's, as the correlations between different PDF parameters
are large.


\begin{thebibliography}{99}
\bibitem{r1} H.~L.~Lai et al., Phys.~~Rev.~ D55 (1997) 1280;\\
             A.~D.~Martin et al., DTP-96-102, Dec 1996, hep-ph/9612449.
\bibitem{r2} The CDF collaboration, F.~ Abe et al.,
             Phys.~Rev.~Lett.~77 (1996) 438.
\bibitem{r3} J.~Huston et al., Phys.~Rev.~Lett.~77 (1996) 444.
\bibitem{r4} The CDF II collaboration, F.~Abe et al., 
             the CDF II detector technical design report, 
             FERMILAB-Pub-96/390-E.
\bibitem{r5a} R.~M.~Barnett et al., Phys.~Rev.~D54, 1 (1996).
\bibitem{r5b} Numerical Recipes, W.~H.~Press et al., 
              Cambridge University Press.
\bibitem{r7} The CDF collaboration, Phys.~Rev.~Lett.~74 (1995) 850;\\
             R.~G.~Wagner, for the CDF Collaboration, 
             Published Proceedings 5th International Conference
             on Physics Beyond the Standard Model, Balholm, Norway, 
             April 29-May 4, 1997.
\bibitem{r6} S.~Alekhin, IFVE-96-79 (Sep 1996), hep-ph/9611213.
\bibitem{CTEQevol} The program was obtained from the 
                   CTEQ collaboration WWW-site 
                   ``http://www.phys.psu.edu/$\sim$cteq/''.
\bibitem{jetrad} W.~T.~Giele, E.~W.~N.~Glover and D.~A.~Kosower, 
                 Nucl.~Phys.~{\bf B403} (1993) 633.
\bibitem{bcdms} The BCDMS collaboration, A.~C.~Benvenuti et al.,
                Phys.~Lett.~223B (1989) 485; Phys.~Lett 237B (1990) 592.
\bibitem{nmc} The NM collaboration, M.~Arneodo et al.,
              Nucl.~Phys.~B483 (1997) 3.
\bibitem{h1} The H1 collaboration, S.~Aid et al., 
             Nucl.~Phys.~470B (1996) 3.
\bibitem{zeus} The ZEUS collaboration, M.~Derrick et al., 
               Zeit.~Phys.~C72 (1996) 399.
\end{thebibliography}
\end{document}